\documentstyle[prl,aps]{revtex} 

\draft

\begin{document}

\twocolumn[\hsize\textwidth\columnwidth\hsize\csname
@twocolumnfalse\endcsname

\title{Quark Coulomb Interactions and the Mass Difference of Mirror Nuclei}

\author{C.~J.~Horowitz
}
\address{Nuclear Theory Center\\ 2401 Milo B.  Sampson Lane \\ Bloomington,
Indiana 47405}

\author{J.  Piekarewicz
}
\address{Department of Physics and Supercomputer Computations Research
Institute, \\ Florida State University, Tallahassee, FL 32306}

\date{\today} 

\maketitle 

\begin{abstract} 

We study the Okamoto-Nolen-Schiffer (ONS) anomaly in the binding
energy of mirror nuclei at high density by adding a single neutron or
proton to a quark gluon plasma. In this high-density limit we find an
anomaly equal to two-thirds of the Coulomb exchange energy of a
proton.  This effect is dominated by quark electromagnetic
interactions---rather than by the up-down quark mass difference. At
normal density we calculate the Coulomb energy of neutron matter
using a string-flip quark model. We find a nonzero Coulomb energy
because of the neutron's charged constituents. This effect could make 
a significant contribution to the ONS anomaly.
\end{abstract} 
\vskip2.0pc]

The Okamoto-Nolen-Schiffer (ONS) anomaly is the long-standing
discrepancy between the calculated and measured binding-energy
differences of mirror nuclei~\cite{Oka64,NolSch69}. The anomaly is
likely to arise from charge symmetry breaking (CSB) in the strong
interaction~\cite{MNS90}, itself believed to originate from
the up-down quark mass difference and electromagnetic effects in the
Standard Model. Thus, the study of CSB is a useful tool to elucidate
the structure of strongly-interacting nuclear systems.

The ONS anomaly can be calculated on several levels. Perhaps the
simplest is the observation by B. A. Brown~\cite{BABro98} that 
the magnitude of the anomaly
is approximately equal to the Coulomb exchange energy. If one adds an
extra proton to a nucleus in a simple Hartree-Fock picture, there will
be both a direct (Hartree) and exchange (Fock) Coulomb interaction
with the other protons. If one---arbitrarily---neglects the Fock term,
one obtains a better agreement with experiment.

At a different level, Blunden and Iqbal compute the ONS anomaly by
calculating the contribution from $\rho$-$\omega$ mixing to the CSB
component of the nucleon-nucleon (NN) interaction~\cite{BluIqb87}.
Their CSB interaction can explain part of the anomaly. However, at
present this is controversial, both in the choice of meson
couplings~\cite{SKB93} and in the momentum dependence of
$\rho$-$\omega$ mixing~\cite{GHT92}. There have been also a number of
calculations of the anomaly based on the contribution from the 
up-down quark mass difference ($\Delta m$). Indeed, Nakamura 
and coworkers~\cite{Nak96} calculate a CSB NN interaction using a
constituent quark model where the short-range color hyperfine
interaction depends explicitly on the quark masses. Moreover, 
the mass difference between the neutron and proton 
may be density dependent~\cite{HK89}. Finally, there 
are other more recent model calculations, such as the one reported
in Ref.~\cite{TST99}.

Although the observation by Brown is not a
dynamical explanation, it is an interesting characterization of the
size of the anomaly. Could there be something wrong with the exchange
term? As the nucleon is a composite object, could it be that the
exchange energy of composite objects yields results significantly
different from the exchange of point nucleons? One expects identical
results if the composite scale of the nucleon is much smaller
than the inter-particle spacing. However these scales are similar in
nuclei.  Moreover, although various calculations based on the up-down
quark mass difference exist, we are not aware of any calculation of
electromagnetic (EM) effects between quarks to the ONS anomaly.

The neutron-proton mass difference in free space is made up 
from comparable contributions of $\Delta m$ and EM effects. Note that 
EM and $\Delta m$ terms contribute with opposite signs to the 
neutron-proton mass difference.  However, the ONS anomaly is 
sensitive to the density dependence of these 
contributions so the relative sign is unknown. In this letter we
study EM effects involving the Coulomb exchange interactions of 
composite nucleons.

To clarify the importance of EM and $\Delta m$ terms we consider a 
high-density limit of the ONS anomaly. We will show, with some mild 
assumptions, that in the high-density limit: (1) there is an ONS 
anomaly and (2) that it is dominated by EM effects with $\Delta m$ 
being unimportant. Further, (3) the magnitude of the anomaly is
simply related to the Coulomb exchange energy and (4) its sign 
is the same as that observed at lower densities. Finally, we will 
perform model calculations to see how relevant this high-density 
limit is to normal-density nuclei.

Consider very high-density symmetric nuclear matter. We assume that
an electron gas makes the system electrically neutral. Thus the direct
Coulomb interaction vanishes. Yet Coulomb exchange effects are still
present. Now add either one proton or one neutron to the system and 
calculate the change in energy. First, model the system as a Fermi gas 
of elementary nucleons. An added proton will have a Coulomb exchange 
energy of
\begin{equation}
  V_p = -e^2 {k_F\over \pi} \;.
 \label{Vproton}
\end{equation}
Here $k_F$ is the Fermi momentum of the proton and $e$ is its electric
charge. In contrast, an added neutron has zero Coulomb exchange energy:
$V_n\!=\!0$. 
Thus, the energy difference between an added
proton and a neutron is just:
\begin{equation}
  E_p - E_n = V_p+ M_p - M_n = -e^2{k_F\over \pi}-\Delta M \;,
 \label{DeltaEH} 
\end{equation}
where $\Delta M=M_n-M_p=1.29$~MeV is the neutron-proton mass
difference. Equation~(\ref{DeltaEH}) is the simple expectation 
of a model with unexcited point nucleons.

Next we consider a quark-gluon plasma. We assume because of 
asymptotic freedom, that at very high density the system is 
nearly a free Fermi gas of quarks. This is because the strong 
coupling $\alpha_S(k_F^2)$ becomes small at the large momentum 
scale characterized by $k_F$ .  When a 
proton is added it will dissociate into two up and one down quark.  
Therefore, the Coulomb exchange energy of these three quarks is 
\begin{equation}
  V_p^{(q)}=-\left(\sum_{i=1}^3 e_i^2\right) {k_F\over \pi} \;,
 \label{Vprotonq}
\end{equation}
where $e_i$ denotes the quark electric charge and $k_F$ is the quark Fermi
momentum.  Note that there are three times as many (valence) quarks as 
nucleons. However the quarks have an extra color degeneracy of three.  
As a result, the quark Fermi momentum in Eq.~(\ref{Vprotonq}) is the
same as the proton's Fermi momentum in Eq.~(\ref{Vproton}). The sum of
the squares of the valence charges in a proton is
$(4/9\!+\!4/9\!+\!1/9)e^2\!=\!e^2$. Because of this ``numerical accident" the
quark Coulomb exchange energy is equal to the Coulomb exchange energy
of an elementary proton.

An interesting difference arises when we add a neutron. In a 
quark-gluon plasma the Coulomb exchange energy is no longer
zero because a neutron is made up of charged constituents. 
Moreover, the exchange energy is always negative independent 
of the sign of the charges; the contributions from positive 
and negative charges add rather than cancel. Indeed, the sum 
of the squares of the valence  quark charges in a neutron is 
$(4/9\!+\!1/9\!+\!1/9)e^2\!=\!2e^2/3$.  
Thus, the neutron Coulomb energy is fully two thirds of that 
of a proton: $V_n^{(q)}\!=\!-{2\over 3}e^2 {k_F/\pi}$. The 
energy difference between an added proton and a neutron becomes:
\begin{equation}
  E_p^{(q)}-E_n^{(q)} = 
  - e^2{k_F\over \pi} + {2\over 3} e^2{k_F\over \pi} =
  - {1\over 3} e^2{k_F\over \pi}\;.
 \label{DeltaEq} 
\end{equation}

We choose to define an ONS anomaly $\Delta E_{\rm ONS}$ as the actual 
energy difference, which we assume is given by Eq.~(\ref{DeltaEq}),
minus the hadronic-model expectation of Eq.~(\ref{DeltaEH}) 
\begin{equation}
 \Delta E_{\rm ONS}\!\equiv\! 
 (E_p^{(q)}\!-\!E_n^{(q)})\!-\!(E_p-E_n)\!=\!
 {2\over 3} e^2 {k_F\over \pi}\!+\!\Delta M \;.
 \label{DeltaEONS} 
\end{equation}
This anomaly arises, not because of an error in the proton's energy
but, because there is a nonzero Coulomb contribution for a
(dissociated) neutron. In principle we should add to the above
equation the contribution from the up-down quark mass difference.
However, in the high-density limit, all contributions from $\Delta m$
are suppressed by the large Fermi momentum. Indeed, the difference in
the Fermi energy of free down and up quarks is:
$\sqrt{k_F^2+m_d^2}\!-\!\sqrt{k_F^2+m_u^2}\!\approx\!(m_d^2-m_u^2)/2k_F$.
Thus, in the limit of very high density the total anomaly---including
contributions from $\Delta m$---becomes dominated by
Eq.~(\ref{DeltaEONS}). Moreover, the original mass difference between 
the neutron and proton ($\Delta M$) ``disappears''at high density
because the contributions from $\Delta m$ are suppressed and the 
Coulomb self-energies of the neutron and the proton are no longer
relevant, as the quarks have rearranged themselves into a uniform 
free Fermi gas.

In summary, we expect that at high density there will be an ONS
anomaly with a magnitude that is two-thirds that of the proton
Coulomb exchange energy. Furthermore, EM effects dominate over the
contribution from $\Delta m$ and the sign of the anomaly is the
same as that observed at normal density.

Our earlier discussion suggests that the Coulomb energy of pure
neutron matter is nonzero. Below we focus on neutron matter because 
of the simple expectation that for point neutrons the Coulomb 
energy is zero. This may provide a signature of substructure.

Since the above statements are only strictly true in the limit of very
high density, we investigate their implications at normal density by
performing a model calculation of neutron matter composed of valence
quarks. While a model is necessary, our philosophy is to use a
``minimal'' one by demanding the following general features that any
realistic model must posses. We require the many-quark wave function
to (1) be explicitly anti-symmetric even for the exchange of quarks
from different nucleons and (2) have cluster separability: the quark
wave function of a nucleon removed to infinity must reduce to that of
a free nucleon, without any unphysical long-range interactions.
Finally, we demand that (3) quarks be confined and (4) for the wave
function to reduce to free nucleons at low density and to a quark
Fermi gas at high density.  Perhaps, any model satisfying these general
features can be used.

Conventional quark potential models with two-body confining
interactions do not satisfy cluster separability as they generate 
unphysical long-range van der Waals interactions between nucleons.  
String-flip models on the other hand, do satisfy the four properties 
described above~\cite{Lenz86,HMN85}.  Unfortunately, we are not aware of
any other models which both satisfy these properties and allow a simple 
calculation.  Thus, we employ the 
three-quark string-flip model discussed in Ref~\cite{HorPie92}.
The model has nonrelativistic constituent quarks of mass $m_c$ of
fixed red, green, and blue colors. A system of $A$ nucleons is modeled 
with $N=3A$ quarks interacting via the following many-body potential:
$V\!=\!V_{RG}\!+\!V_{GB}\!+\!V_{BR}$, where each term represents the
optimal pairing of quarks. For example, the ``red-green'' component
of the potential is defined as
\begin{equation}
  V_{RG}={\rm Min} \left\{\sum_{j=1}^A v({\bf r}_j^{(R)}-
     {\bf r}_{P_j}^{(G)})\right\} \;.
 \label{VRG}
\end{equation}
Here ${\bf r}_j^{(R)}$ is the coordinate of the $j_{th}$ red quark 
and ${\bf r}_{P_j}^{(G)}$ is its green partner in the neutron. The 
minimum is over all $A!$ permutations $P_j$ of the set of $A$ green 
quarks. A harmonic string potential $v(r)=kr^2/2$ is used to confine 
the quarks and the Hamiltonian for the model becomes
\begin{equation}
  H=\sum_{i=1}^{N}\frac{{\bf P}_{i}^{2}}{2m_c}+V 
   =-\sum_{i=1}^N \frac{\nabla_i^2}{2m_c}+V \;.
 \label{Hamiltonian}
\end{equation}
Each red quark is connected by harmonic strings to one and only one
green and to one and only one blue quark. This insures that quarks
will be confined into ``color-neutral'' clusters. For three quarks the
model reduces to the well-known harmonic oscillator quark model.  For
neutron matter there is a very large number of permutations or ways to
connect the strings. We employ an implementation of the linear sum
assignment algorithm by Burkard and Derigs that efficiently finds the
optimal permutation in a time proportional to
$N^3$~\cite{BurDer80}. This allows Monte Carlo simulations with
hundreds of quarks.

The model has two dimension-full parameters: $k$ and $m_c$. Yet we are
only interested in the harmonic-oscillator length $b=(km_c)^{-1/4}$,
as this sets the length scale for quark confinement. The root mean 
square radius of a nucleon is $\langle r^2\rangle^{1/2}=3^{-1/4}b$. 
Hence, to reproduce the experimental charge radius of the proton 
$\langle r^2\rangle^{1/2}=0.86$~fm we choose $b=1.13$~fm. At the
end we can rescale our results for other values of $b$.

We are interested in simulating neutron matter. Therefore we assign 
to red and green quarks an electromagnetic charge of $-e/3$ and to
blue quarks a charge of $2e/3$. For simplicity we do not include any 
other intrinsic degree of freedom, such as spin or isospin. The 
electromagnetic self-energy of an isolated neutron is 
($\alpha\!=\!e^2\!=\!1/137$)
\begin{equation}
  V_n^0=-\sqrt{\frac{2}{9\pi}}
  \frac{\alpha}{\langle r^2\rangle^{1/2}}=-0.446~{\rm MeV} \;.
 \label{vnself}
\end{equation}
A simple variational wave function for the many-quark system has
been constructed in Ref.~\cite{HorPie92}. It is given by 
\begin{equation}
 \Psi = \exp\left(-\lambda {V\over kb^2}\right) \Phi \;,
 \label{Psivar}
\end{equation}
with $\Phi$ a product of Slater determinants for the red, green, 
and blue quarks. In Ref.~\cite{HorPie92} $\lambda$ is a variational 
parameter characterizing the length scale for quark confinement. At 
low density a value of $\lambda\!=\!1/\sqrt{3}$ allows
Eq.~(\ref{Psivar}) to reproduce the gaussian wave function of a free
nucleon. For simplicity we keep lambda fixed at
$\lambda\!=\!1/\sqrt{3}$ for all densities. This insures that any change 
in the Coulomb energy of a neutron does not arise from an artificial 
change in this length scale.


We calculate the total Coulomb energy
\begin{equation}
  V_{\rm Coul}^{\rm tot}=\sum_{i<j}^N 
  \frac{e_ie_j}{|{\bf r}_i-{\bf r}_j|} \;,
 \label{VCoulomb}
\end{equation}
of a system of $N\!=\!3A$ quarks in a box of volume $V$ with
antiperiodic boundary conditions. To minimize finite size effects we
use periodic distances to compute the quark separation. The neutron
density of the system is $\rho_n\!=\!{A/V}$. We use standard
Metropolis Monte Carlo techniques to calculate the expectation value
of the total Coulomb energy for the wave function given in
Eq.~(\ref{Psivar}).

Figure 1 shows the change in the Coulomb energy per neutron 
\begin{equation}
  \Delta V\equiv 
  {1\over A}\langle V_{\rm Coul}^{\rm tot}\rangle - V_n^0 \;,
 \label{DeltaV}
\end{equation}
as a function of density for systems with $N$=96 and 264 quarks.  
We have subtracted the neutron self-energy $V_n^0$ of 
Eq.~(\ref{vnself}) because this is included in the experimental 
neutron-proton mass difference. We find $\Delta V$ to be nonzero.

\vbox to 2.6in{\vss\hbox to 8in{\hss {\includegraphics{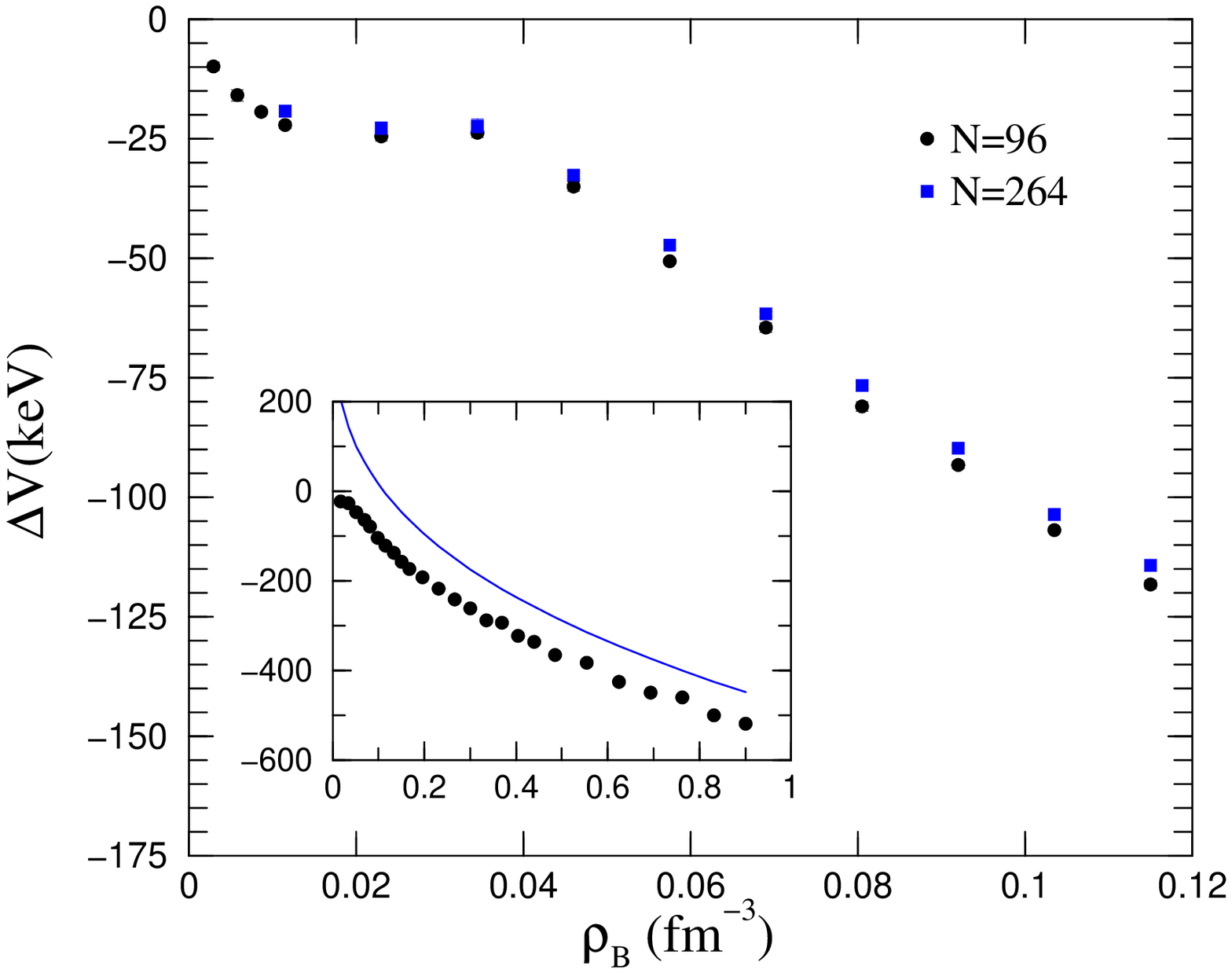}}\hss}} \nobreak
{\noindent\narrower{{\bf FIG.~1}. Change in Coulomb energy per neutron 
as a function of baryon density for pure neutron matter. The insert 
compares the model to a free Fermi gas (solid line) at high density.}}
\vskip 10pt

At normal density $\rho_n\!=\!0.08$~fm$^{-3}$ and $N=96$:
$\Delta V = -78 \pm 1~{\rm keV}$.
The scale of this result suggests that changes in the 
Coulomb energies of quarks can make a significant contribution 
to the ONS anomaly. More refined models may give results which are 
of the same order of magnitude, given the ratio of the 
nucleon size to interparticle spacing.  Furthermore, we 
expect an additional contribution from the up-down quark mass 
difference $\Delta m$. Our result is 
somewhat smaller than the total observed anomaly of the order 
200 keV in mass 15 and 300 keV in mass 39~\cite{MHSha94}. Note that, 
for simplicity, we have calculated the average Coulomb energy 
per neutron rather than the self-energy of a single valence neutron.  
These quantities are expected to be similar. Indeed, in a free 
Fermi gas the average Coulomb energy per proton is three fourths 
of that of Eq.~(\ref{Vproton}).

Figure 2 shows $\Delta V$ as a function of the nucleon 
root mean square radius or oscillator length at the fixed density of
$\rho_n=0.08$~fm$^{-3}$. Making the quark core of a nucleon smaller 
reduces $\Delta V$, but not by much. Further, as the oscillator
length is made very small the scale of the neutron self-energy 
$V_n^0$ grows and this can increase $\Delta V$. Of course, if 
the nucleon core is small one must use a large meson cloud to 
account for the full proton charge radius. This meson cloud, 
which we have not included, could further increase $\Delta V$.

\vbox to 2.6in{\vss\hbox to 8in{\hss {\includegraphics{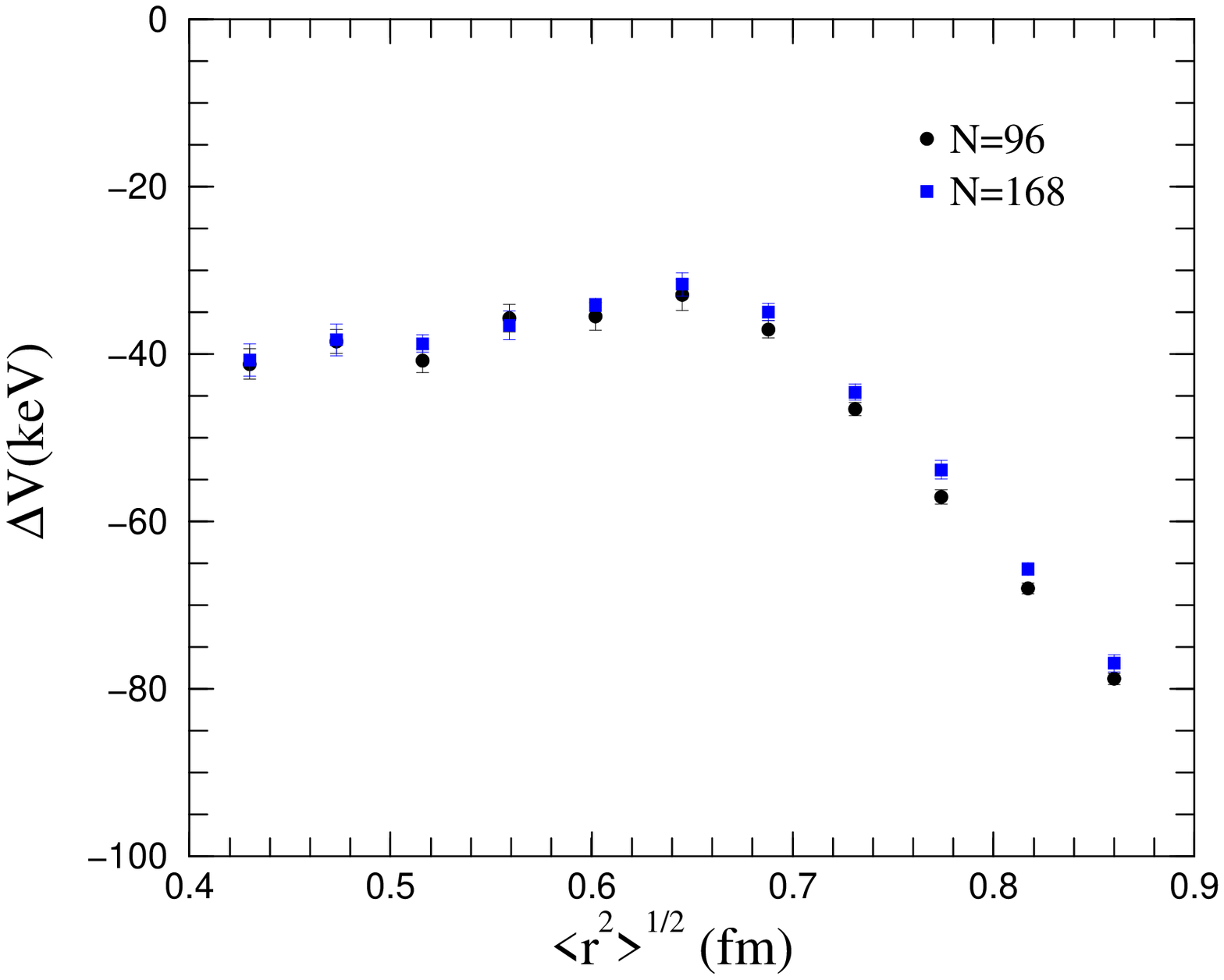}}\hss}} \nobreak
{\noindent\narrower{{\bf FIG.~2}. Change in the Coulomb energy per 
neutron as a function of the nucleon root mean square radius 
at the fixed neutron density of $\rho_n\!=\!0.08$~fm$^{-3}$.}}
\vskip 10pt

One should extend our results by using more elaborated quark models.
It is important to study models with more intrinsic spin and flavor 
degrees of freedom along with more complete treatments of color.  
However, in these more complete models we still expect an exchange or
dynamical correlation between quarks associated with the nucleon's 
hard core.  This correlation could lead to a nonzero Coulomb energy 
for neutrons.  Note that we have used harmonic oscillator confining 
strings. Thus, our wave function has gaussian tails. Linear confinement 
may increase the tails and this should enhance the Coulomb exchange energy.

In conclusion, we have considered a high-density limit of the
Okamoto-Nolen-Schiffer anomaly to clarify the role of electromagnetic
interactions (EM) and of the up-down quark mass difference $\Delta
m$. We have added a single neutron or proton to a quark gluon
plasma. In this high-density limit we find that: (1) there is an ONS
anomaly, (2) it is dominated by EM interactions rather than by $\Delta
m$, and (3) its magnitude is two-thirds of the proton Coulomb exchange
energy. We find an attractive Coulomb exchange energy for an added
neutron because of the neutron's charged constituents. This suggests
that the ONS anomaly could be closely related to the nucleon
substructure.  We use a minimal string-flip quark model to calculate
the Coulomb energy of pure neutron matter. The model wave function is
fully anti-symmetric and satisfies cluster separability and quark
confinement.  At normal density, we find a nonzero Coulomb energy 
for neutron matter that could make a significant contribution to 
the ONS anomaly.

This work was supported in part by DOE grants DE-FG02-87ER40365,
DE-FC05-85ER250000, and DE-FG05-92ER40750.


\begin{references}
\bibitem{Oka64}    K. Okamoto, Phys. Lett. {\bf 11} (1964) 150.
\bibitem{NolSch69} J. A. Nolen and J. P. Schiffer, Ann. Rev. 
                   Nucl. Sci. {\bf 19} (1969) 471.
\bibitem{MNS90}    G. A. Miller, B. M. K. Nefkens and I. Slaus, 
	           Phys. Rep. {\bf 194} (1990) 1.
\bibitem{BABro98}  B. Alex Brown, Phys. Rev. {\bf C58} (1998) 220.
\bibitem{BluIqb87} P. G. Blunden and M. J. Iqbal, 
	           Phys. Lett. {\bf B198} (1987) 14.
\bibitem{SKB93}    Thomas Schafer, Volker Koch and Gerald E. Brown, 
	           Nuc. Phys. {\bf A562} (1993) 644.
\bibitem{GHT92}    T. Goldman, J. A. Henderson and A. W. Thomas, 
	           Few-Body Systems, {\bf 12} (1992) 123.
\bibitem{Nak96}    S. Nakamura, K. Muto, M. Oka, S. Takeuchi and
		   T. Oda,  Phys. Rev. Lett. {\bf 76} (1996) 881.
\bibitem{HK89}     E. M. Henley and G. Krein, Phys. Rev. Lett. {\bf 62} (1989)
                   2586.
\bibitem{TST99}    K. Tsushima, K. Saito and A. W. Thomas, 
	           Phys. Lett. {\bf B465} (1999) 36.
\bibitem{Lenz86}   F. Lenz et al., Ann. of Phys. (N.Y.) {\bf 170} (1986) 65.
\bibitem{HMN85}    C. J. Horowitz, E. Moniz and J. Negele, Phys. Rev. {\bf D31} (1985) 1689.
\bibitem{HorPie92} C. J. Horowitz and J. Piekarewicz, Nuc. Phys. {\bf A536} (1992) 669.
\bibitem{BurDer80} R.E.~Burkard and U.~Derigs, 
	           {\it Lecture Notes in Economics and Math Systems}, 
	           {\bf 184} (Springer-Verlag) 1980, Chapter~1.
\bibitem{MHSha94}  M. H. Shahnas, Phys. Rev. {\bf C50} (1994) 2346.
\end{references}
\end{document}